# Inverse orbital Hall effect induced terahertz emission enabled by a ferromagnet with quenched orbital moment in Fe/Pt/W trilayers

Chao Zhou [1†], Lei Hao[1†], Shaohua Zhang[1†], Yaxuan Jin[1], Xianguo Jiang[1],
Ning Yang[1], Li Zheng[1], Hao Meng[1], Chao Lu[1], Wendeng Huang[1],
Yizheng Wu[2,3*], Yan Zhou[4*], Jia Xu[1*]

[1.] Department of Physics, School of Physics and Telecommunication Engineering, Shaanxi University of Technology, Hanzhong 723001, China

[2.] Department of Physics and State Key Laboratory of Surface Physics, Fudan University, Shanghai 200433, China

[3.] Shanghai Research Center for Quantum Sciences, Shanghai 201315, China

[4.] Guangdong Basic Research Center of Excellence for Aggregate Science, School of Science and Engineering, The Chinese University of Hong Kong, Shenzhen, Shenzhen, Guangdong 518172, China

†These authors contributed equally to this work.

## Abstract

The inverse orbital Hall effect (IOHE) has recently attracted considerable attention as an emerging mechanism for terahertz (THz) emission based on ultrafast angular-momentum-to-charge conversion. Most experimental studies have focused on materials with strong spin-orbit coupling or pronounced orbital character, where sizable orbital Hall responses are expected. Elemental ferromagnets such as Fe are generally regarded as quenched orbital sources and are not expected to exhibit orbital-dominated THz emission. Here, we report a pronounced enhancement of THz emission in Fe/Pt/W trilayer heterostructures, despite the absence of detectable orbital contributions in the corresponding Fe/Pt and Fe/W bilayers. Thickness-dependent measurements reveal long-distance signal persistence, systematic delay accumulation, and pronounced pulse broadening with increasing W thickness. These features are inconsistent with diffusive spin transport and indicate that orbital angular momentum transport in the W layer, converted into charge current via the IOHE, becomes a dominant channel for THz emission in the trilayer configuration. Our results demonstrate that strong IOHE can emerge in heterostructures incorporating a quenched orbital ferromagnet, providing an effective route to enhance spintronic THz emitters through orbital Hall physics.

# 1. Introduction

Spintronic terahertz (THz) emitters based on ultrafast angular momentum-to-charge conversion have emerged as a versatile platform for broadband THz generation, combining sub-picosecond temporal resolution, compact geometry, and compatibility with on-chip integration [1-3]. In conventional ferromagnet (FM)/heavy-metal (HM) heterostructures, femtosecond laser excitation generates nonequilibrium spin currents in the FM layer, which are converted into transverse charge currents via the inverse spin Hall effect (ISHE), giving rise to single-cycle THz emission [4-7]. This ISHE-based mechanism has been widely demonstrated in various FM/HM systems [8-11].

A fundamental limitation of ISHE-driven THz emitters, however, is the short spin diffusion length in typical heavy metals such as Pt and W [12, 13], which restricts efficient charge conversion to ultrathin layers. As a result, increasing the HM thickness leads to rapid signal attenuation due to spin relaxation and THz absorption. This constraint has stimulated growing interest in alternative transport channels that could enable long-range angular momentum propagation.

Recently, orbital angular momentum has been recognized as a promising complement to spin transport [14-19]. Both theoretical and experimental studies have shown that orbital currents generated via the orbital Hall effect (OHE) can propagate over much longer distances and be converted into charge currents through inverse orbital conversion mechanisms such as the inverse orbital Hall effect (IOHE) or inverse orbital Rashba-Edelstein effect (IOREE) [20-22]. These orbital-mediated processes have been linked to distinct THz emission signatures, including long-distance signal persistence, thickness-dependent delay accumulation, and pulse broadening [23-26].

Despite these advances, orbital-dominated THz emission has so far been demonstrated mainly in systems specifically engineered to support strong orbital responses [27, 28]. Recent studies have reported enhanced THz emission in CoPt/Pt/W heterostructures, where orbital angular momentum plays an important role [29]. In such systems, the ferromagnetic CoPt alloy itself acts as a strong intrinsic source of orbital currents, and both Pt and W possess orbital Hall angles of the same sign, facilitating constructive orbital contributions. In contrast, elemental ferromagnets such as Fe are generally regarded as having their orbital moment quenched, and IOHE-related contributions are therefore expected to be negligible in Fe-based heterostructures [30]. Consistent with this expectation, Fe/Pt and Fe/W bilayers typically exhibit purely spin-current-dominated THz emission, with no clear signatures of long-range orbital angular momentum transport [31, 32].

In this work, we demonstrate that this expectation breaks down in Fe/Pt/W trilayers. Although both Fe/Pt and Fe/W bilayers are purely ISHE-dominated, the Fe/Pt/W trilayer exhibits a pronounced enhancement of THz emission accompanied by long-distance signal persistence, systematic delay accumulation, and pulse broadening. These features cannot be explained by diffusive spin transport alone. By systematically varying the Pt and W thicknesses, we demonstrate that orbital angular momentum transport in the W layer, converted into charge current via the IOHE, becomes a dominant channel for THz emission in Fe/Pt/W trilayers. Importantly, this strong orbital contribution emerges only in the trilayer configuration and is absent in the corresponding bilayers, despite the quenched orbital character of Fe. Our results establish a clear experimental route to activate orbital Hall physics in heterostructures based on elemental ferromagnets and highlight trilayer design as an effective strategy for enhancing spintronic THz emitters.

## 2. Experiments

**Sample Fabrication:**

A series of Fe/Pt/W heterostructures were fabricated on 1-mm-thick glass substrates using magnetron sputtering at room temperature. The heavy-metal layers of Pt and W were deposited by DC sputtering, while the Fe layer was grown by RF sputtering. The base pressure of the chamber was maintained below $3.5\times10^{-5}$ Pa, and working pressure during deposition was approximately 0.2 Pa. The deposition rates were calibrated to about 1 nm/min for Pt, and about 2 nm/min for Fe and W. The layer thickness was controlled by deposition time, ranging from 0 to 100 nm depending on the designed structure. All stacking sequences are described from bottom to top, with the first layer deposited directly onto the glass substrate. The thickness of the ferromagnetic Fe layer was fixed at 2 nm for all samples to ensure consistent angular momentum injection characteristics.

**THz Emission Measurements:** THz emission spectroscopy measurements were performed at room temperature in a dry air environment with relative humidity below 5%. A schematic diagram of the THz emission measurement setup is presented in Figure 1a. During measurements, samples were mounted in an electromagnet providing a 1000 Oe in-plane magnetic field along the x-axis, with the laser pulses incident normally from the substrate side. The laser pulses used in the experiments are linearly polarized, which are generated from a Ti:Sapphire laser oscillator (with a duration of ~80 fs, a center wavelength of 800 nm, and a repetition rate of 80 MHz). The pulse energy of the laser beam was kept below ~1.7 nJ to ensure that the emitted THz signal was linearly proportional to the laser intensity. The laser beam diameter was approximately 100 μm. The emitted THz signal was detected using the electro-optic sampling technique with a 0.5 mm-thick electro-optic ZnTe(110) crystal.

## 3. Results and discussion
### A. Emergent THz emission enhancement in Fe/Pt/W trilayers

Figure 1 presents a comparative overview of the THz emission characteristics of Fe/Pt and Fe/W bilayers and the Fe/Pt/W trilayer. As schematically illustrated in Figs. 1(a) and (b), both Fe/Pt and Fe/W bilayers are expected to exhibit THz emission dominated by ISHE. Owing to the opposite signs of the spin Hall angles of Pt ($\theta_{SH,Pt} > 0$) and W ($\theta_{SH,W} < 0$) [32-35], the two bilayers generate THz waveforms with opposite polarities, as confirmed by the time-domain signals measured from Fe(2 nm)/Pt(1.5 nm) and Fe(2 nm)/W(0.75 nm) in Fig. 1(d). Notably, the THz amplitude of the Fe/Pt bilayer is significantly larger than that of Fe/W, which can be attributed to the larger effective spin Hall angle and thus, higher spin-to-charge conversion efficiency of Pt [36].

Based on these bilayer responses, a simple superposition model would predict that in a Fe/Pt/W trilayer the spin currents injected into Pt and W should generate THz emissions of opposite polarity, leading to partial cancellation and therefore a reduced overall THz output. Contrary to this expectation, the Fe/Pt/W trilayer exhibits a pronounced enhancement of the THz emission. As shown by the red curve in Fig. 1(d), the peak-to-peak amplitude $\Delta V$ of the trilayer far exceeds those of both Fe/Pt and Fe/W, and this non-additive enhancement is summarized in Fig. 1(e).

This anomalous enhancement immediately indicates that the THz response of the Fe/Pt/W trilayer cannot be explained by conventional spin-current-driven ISHE alone. Instead, the trilayer geometry enables an additional ultrafast angular momentum conversion pathway. As illustrated in Fig. 1(c), the Pt layer can enable conversion of spin angular momentum into orbital angular momentum, through its strong spin-orbit coupling [37]. The orbital currents might propagate over longer distances and be injected into the W layer, where they are converted into transverse charge currents via IOHE [29]. The constructive contribution of this orbital-to-charge conversion channel could then lead to the strongly enhanced THz emission observed in the Fe/Pt/W trilayer.

**B. Spin-current-dominated THz emission in Fe/Pt and Fe/W bilayers**

To establish a reliable baseline for spin-current-driven THz emission and to clarify whether orbital transport plays a role in Fe-based bilayers, systematic control experiments were performed on Fe/Pt and Fe/W bilayers with varying heavy-metal thicknesses. Figures 2(a) and 2(b) show representative time-domain THz waveforms of Fe/Pt and Fe/W bilayers at different thicknesses. In both systems, the THz amplitude decreases rapidly as the HM thickness increases.

The extracted peak-to-peak THz amplitudes $\Delta V$ are summarized in Fig. 2(c). For Fe(2 nm)/Pt($d_{Pt}$), $\Delta V$ reaches a maximum at $d_{Pt} \approx 2$ nm and then decreases sharply, approaching nearly zero at $d_{Pt} \geq 15$ nm. A similar thickness-dependent attenuation is observed in Fe(2 nm)/W($d_W$), although with opposite polarity due to the negative spin Hall angle of W. Such rapid decay is characteristic of ISHE-dominated charge conversion occurring within the initial few nanometers of the heavy-metal layer, together with enhanced THz absorption and electrical screening in thicker metallic films.

Despite the strong amplitude attenuation, the temporal position of the THz peak remains nearly unchanged across the entire thickness range for both bilayers. As shown in Fig. 2(d), the extracted delay time $\tau_D$, obtained from the THz waveform (see details in Fig. S1 of the Supplementary Information [38]), varies only weakly with heavy-metal thickness and remains within a few femtoseconds of the reference value. This thickness-independent temporal behavior demonstrates that the THz pulse arrival time is governed by short-range diffusive spin transport rather than by long-range angular momentum propagation.

Further insight is provided by the THz pulse width analysis shown in Fig. 2(e). The pulse width $\Delta t$ exhibits no systematic dependence on heavy-metal thickness for either Fe/Pt or Fe/W bilayers. Moreover, normalized time-domain waveforms at different thicknesses collapse onto a single curve (see Supplementary Fig. S2), indicating the absence of dispersive transport effects. These results unambiguously establish that THz emission in Fe/Pt and Fe/W bilayers is dominated by spin-current diffusion and ISHE-based charge conversion, with negligible contribution from orbital Hall transport [32]. This behavior is consistent with the quenched orbital moment of Fe, which is generally expected to generate only a small orbital angular momentum component [14].

**C. Long-range angular momentum transport in Fe/Pt/W trilayers**

In sharp contrast to the bilayer systems, Fe/Pt/W trilayers exhibit clear signatures of long-range angular momentum transport. Figure 3(a) shows representative time-domain THz waveforms of Fe(2 nm)/Pt(2 nm)/W($d_W$) as the W thickness is increased from 1 nm to 100 nm. Although the THz amplitude decreases monotonically with increasing $d_W$, a sizable THz signal remains detectable even at $d_W$ = 100 nm. Such long-distance persistence of the THz signal is incompatible with spin diffusion, whose characteristic length in W is only a few nanometers [34].

The W thickness dependence of the signal amplitude $\Delta V$ for different Pt thicknesses is summarized in Fig. 3(b). Above the thickness of a few nanometers, while $\Delta V$ decreases with increasing $d_W$ in all cases, the decay is significantly slower than that observed in Fe/W bilayers. To quantify this behavior, the dependence of $\Delta V$ on the W thickness in the regime of $d_W > 10$ nm is fitted using an exponential formula:

$$\Delta V(d_W) = \Delta V_0 \exp(-d_W/d_h) \tag{1}$$

where $\Delta V_0$ is the THz signal amplitude and $d_h$ is an effective decay length. As shown in Fig. 3(c), $d_h$ exhibits a pronounced dependence on Pt thickness and reaches a maximum near $d_{Pt} \approx 2$ nm, indicating the slowest rate of decay.

The THz pulse width exhibits pronounced broadening with increasing W thickness when the Pt thickness is fixed at 2 nm, as shown in Fig. 3(d). In contrast, for Pt thicknesses of 1 nm and 3 nm the pulse width shows much weaker dependence on W thickness. Temporal analysis provides direct evidence for propagating angular momentum. As shown in Fig. 3(e), the delay time $\tau_D$ increases approximately linearly with W thickness for all investigated Pt thicknesses. Assuming a ballistic transport regime, the transport velocity can be found as the inverse of slope, as in Fig. 3(f). The linear fits yield effective propagation velocities on the order of 0.3~0.6 nm/fs for Fe/Pt/W trilayer, and ~1.0 nm/fs for Fe/W bilayer. The high velocity observed in Fe/W agrees with the properties of ultrafast spin currents [39]. The significantly slower transport in Fe/Pt/W is inconsistent with spin diffusion, and suggested orbital angular momentum propagation in the W layer [40]. The slowest velocity occurred when Pt is 2 nm, suggesting that the contribution from IOHE might be strongest in this case. The coexistence of long-distance propagation, thickness-dependent delay accumulation, and pulse broadening provides compelling evidence that orbital Hall transport and IOHE-based charge conversion dominate the THz emission process in Fe/Pt/W trilayers.

### D. Thickness-matched ISHE-IOHE cooperative enhancement mechanism

The thickness dependence of the THz enhancement reveals that orbital transport alone is not sufficient to maximize the THz output. Figure 4(a) depicts the mechanism for THz emission in Fe/Pt/W: upon laser excitation, a strong spin current $j_S$ and a quenched orbital current $j_L$ are both injected into Pt. Through the ISHE in Pt, $j_S$ is converted into a charge current $j_C$, and generates THz emission. On the other hand, the strong spin orbit coupling in Pt also converts $j_S$ into $j_L$ [29, 37]. The orbital current from the conversion in Pt can be much stronger than that from Fe directly. In total, $j_L$ gets converted into $j_C$ via the IOHE in W, also creating THz emission, the charge currents generated by the ISHE in Pt and the IOHE in W possess the same polarity, allowing the spin- and orbital-originated THz signals to coherently interfere and jointly modulate the overall THz emission amplitude. To learn about the cooperation of THz emission from both mechanisms, we investigated the evolution of THz emission when the thickness of W is very thin.

Figure 4(b) shows the evolution of the THz signal from Fe(2 nm)/Pt($d_{Pt}$)/W($d_W$) when the W thickness $d_W$ is fixed at 0.75 nm. As $d_{Pt}$ increases from 0 to 2 nm, the THz amplitude increases monotonically, indicating a progressively enhanced emission. When the $d_{Pt}$ exceeds 2 nm, however, the THz signal gradually decreases. In contrast, as shown in Fig. 4(c), when the $d_{Pt}$ is fixed at 2 nm, the THz signal exhibits a different dependence on the $d_W$, reaching its maximum at $d_W \approx 0.5$ nm. This distinct behavior highlights that the optimal enhancement condition depends sensitively on the thickness matching between Pt and W layers.

As shown in Fig. 4(d), when the thickness of W is fixed at 0.75 nm and 2 nm, $\Delta V$ initially always increases with $d_{Pt}$, and a maximum value is reached when $d_{Pt}$ is about 2 nm. To better describe the enhancement from the cooperation of ISHE and IOHE, we have defined an enhancement factor $\eta$:

$$\eta(d_{Pt}, d_W) = \frac{\Delta V_{Fe/Pt(d_{Pt})/W(d_W)}}{\Delta V_{Fe/Pt(d_{Pt})} + \Delta V_{Fe/W(d_W)}} \tag{2}$$

The trends for $\eta$ are summarized in Fig. 4(e). It is clear that when $d_W$ is 0.75 nm, there is a wide range of $d_{Pt}$ (0.5~2.5 nm) that enables enhancement of THz emission, with enhancement factor $\eta>1$. When $d_W$ is increased to 2 nm, the value of $\eta$ becomes generally smaller than 1, with only one exception at around $d_{Pt}$ =1~1.5 nm. Those results suggest $d_{Pt}$ ~1.5 nm as the condition to achieve optimal enhancement.

On the other hand, the trends in $\Delta V$ as a function of $d_W$ are shown in Fig. 4(f). It is clear that as $d_{Pt}$ increases, the corresponding $d_W$ of peak gradually shifts to the left until it becomes zero, leading to monotonically decreasing signal when Pt thickness is 3 nm. The trends of $\eta$ for the corresponding samples are summarized in Fig. 4(g), revealing its complex dependence on $d_W$.

When $d_{Pt}$ is 1 nm, $\eta$ first increases with $d_W$, reaching a peak at 2 nm, and then decreases. At around $d_W$~4 nm, $\eta$ passes the threshold of 1, suggesting a conversion from the enhancement region to a surppression one. As $d_{Pt}$ increases to 2 nm, the peak for $\eta$ occurred at $d_W$ ~0.5 nm.

This behavior reflects a thickness-matched cooperative mechanism between ISHE and IOHE. For ultrathin Pt layers, partial orbital injection into W can already occur, giving rise to detectable IOHE-related signatures. However, the Pt layer might not be fully continuous and the ISHE-induced charge current is spatially inhomogeneous, limiting the cooperative enhancement. When the Pt thickness increases to approximately 2 nm, the spin current injected from Fe is more efficiently converted into a transverse charge current via ISHE in Pt, while simultaneously generating a strong orbital polarization through spin-orbit coupling. The resulting orbital current is subsequently injected into W and converted into an additional charge current via IOHE. The temporal and spatial overlap between the prompt ISHE response in Pt and the delayed IOHE contribution in W leads to the maximum THz enhancement.

For Pt layers thicker than 3nm, orbital transport and IOHE conversion in W remain present, as evidenced by the continued delay accumulation and pulse broadening [15]. However, the spin current injected from Fe is largely relaxed within the Pt layer before reaching the Pt/W interface, resulting in a reduced ISHE contribution. Consequently, the cooperative ISHE-IOHE mechanism becomes inefficient, and no further enhancement of the THz amplitude is observed despite the persistence of orbital transport.

In addition, reversing the stacking order of Pt and W would disable both the emission enhancement and the long-distance transport (See Supplementary Figure S5). That highlights the importance for the strong spin-to-orbital conversion in Pt, along with the strong IOHE in W in promoting such effects. Overall, these results demonstrate that the enhanced THz emission in Fe/Pt/W trilayers originates from a thickness-matched ISHE-IOHE cooperative mechanism, in which the Pt/W combination effectively amplifies the otherwise quenched orbital character of Fe. This work highlights the importance of interface engineering and thickness optimization in activating orbital angular momentum transport and provides a practical strategy for designing high-performance spin-orbitronic THz emitters.

## 4. Conclusion

In conclusion, we have demonstrated a pronounced and non-additive enhancement of THz emission in Fe/Pt/W trilayers that cannot be explained by conventional ISHE alone. Through systematic thickness-dependent and stacking-sequence-dependent investigations, we establish that Fe/Pt and Fe/W bilayers serve as clean spin-dominated reference systems, consistent with quenched orbital character of Fe. In these bilayers, THz emission is governed by short-range spin diffusion and ISHE-based charge conversion, with no detectable signatures of orbital Hall transport.

In sharp contrast, the Fe/Pt/W trilayer exhibits clear evidence of long-range angular momentum transport, including systematic delay accumulation, pulse broadening, and finite THz signals persisting over tens of nanometers in W. These features unambiguously indicate the activation of orbital angular momentum transport and IOHE-based charge conversion. Crucially, the observed THz enhancement arises from a thickness-matched cooperative mechanism between ISHE in Pt and IOHE in W. When the Pt thickness is optimized (~2 nm), spin-to-charge conversion in Pt and orbital-to-charge conversion in W act constructively in both space and time, leading to maximal THz emission. For thinner or thicker Pt layers, although orbital transport and IOHE signatures may still exist, the cooperative enhancement is suppressed due to incomplete film continuity or excessive spin relaxation.

Our results establish ISHE-IOHE cooperation as a powerful and tunable route to enhance ultrafast THz emission, and demonstrate that orbital angular momentum, though quenched in conventional ferromagnets, can be effectively activated and amplified by interface engineering and thickness optimization. Beyond THz emitters, our trilayer strategy suggests a viable pathway to harness orbital degrees of freedom using standard ferromagnetic materials, opening opportunities for orbital-assisted information storage and logic functionalities in ultrafast spin–orbitronic devices.

**Acknowledgements:**

This work was supported by the National Natural Science Foundation of China (Grant No. 12204295 and 12204296), the Open Research Project of State Key Laboratory of Surface Physics from Fudan University (Grant No. KF2022_17), the Youth Science and Technology Star Program of Shaanxi Province (2025ZC-KJXX-15), the Young Talent Fund of Association for Science and Technology in Shaanxi (20250512), the Youth Project in Sanqin Talent Introduction Program, and the Shaanxi Province Outstanding Young Talent Support Program for Universities and Young Hanjiang Scholar of Shaanxi University of Technology, and Shaanxi University of Technology (Grants No. SLGRCQD044 and SLGRCQD046). The Innovation Fund of Shaanxi University of Technology (Grant No. SLGYCX2514 and SLGYCX2511). Yizheng Wu acknowledges the support from the National Key Research and Development Program of China (Grant No. 2024YFA1408501), the National Natural Science Foundation of China (Grant No.11974079, 12274083, 1222100), the Shanghai Municipal Science and Technology Major Project (Grant No. 2019SHZDZX01), and the Shanghai Municipal Science and Technology Basic Research Project (Grant No. 22JC1400200). Hao Meng acknowledges the support from the National Natural Science Foundation of China (Grant No.12174238). Y. Z. acknowledges support by the Shenzhen Fundamental Research Fund (Grant No. JCYJ20210324120213037), and the National Natural Science Foundation of China (12374123), Guangdong Basic Research Center of Excellence for Aggregate Science, and the 2023 SZSTI stable support scheme.

**Figures:**

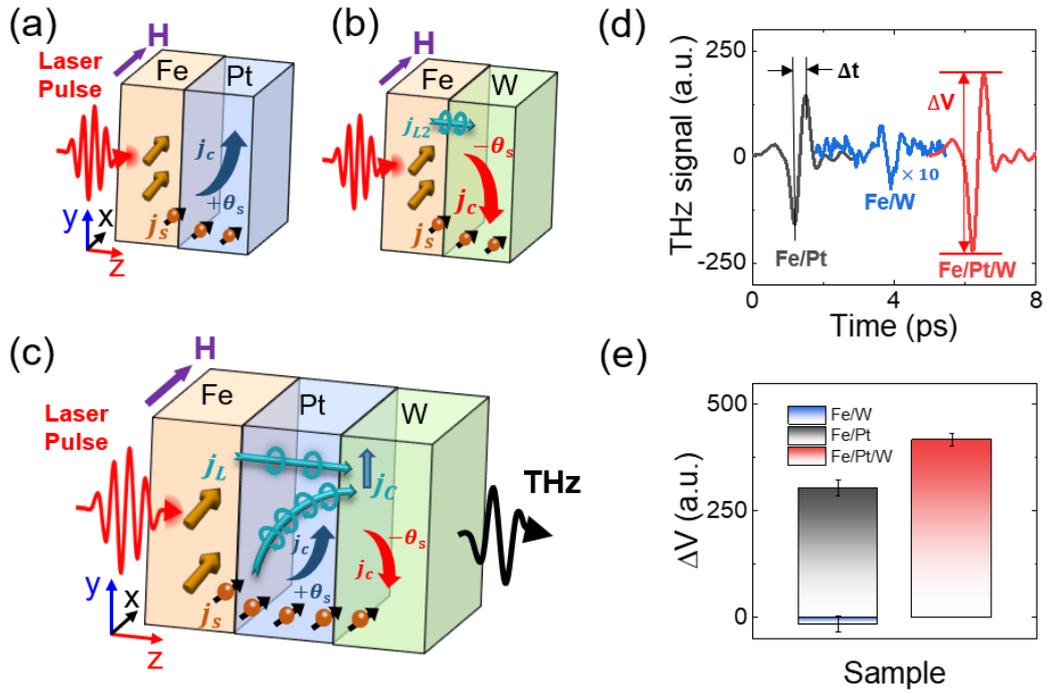

**Figure 1** (a)-(c) Schematic illustrations of (a) Fe/Pt, (b) Fe/W, and (c) Fe/Pt/W bilayers under femtosecond laser excitation. (d) Representative time-domain THz waveforms emitted from Fe/Pt, Fe/W, and Fe/Pt/W samples, highlighting the enhanced peak-to-peak amplitude $\Delta V$ in the trilayer. The curves are shifted horizontally for clarity. (e) Comparison of the extracted $\Delta V$ for different sample structures in (d).

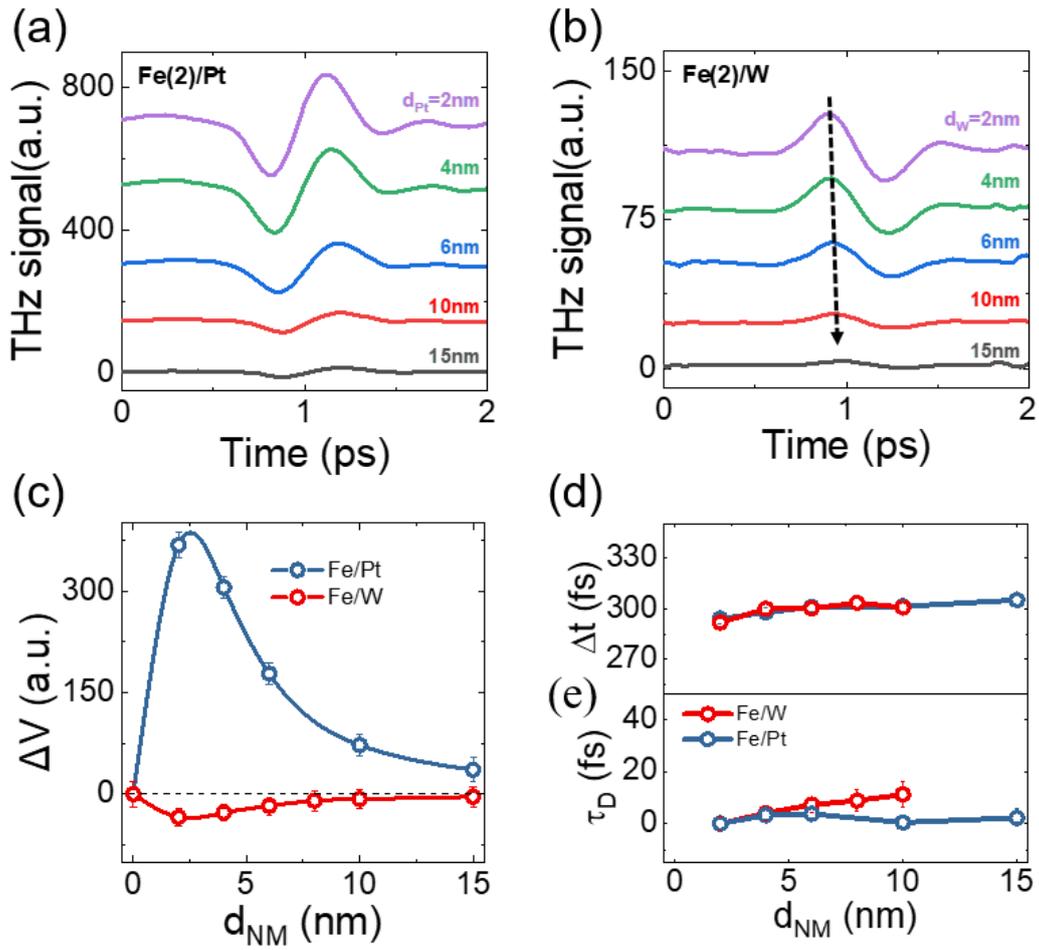

**Figure 2** (a)-(b) Time-domain THz waveforms emitted from Fe(2 nm)/Pt($d_{Pt}$) and Fe(2 nm)/W($d_W$) bilayers with different nonmagnetic layer thicknesses, respectively. (c) Extracted peak-to-peak THz amplitudes ΔV as a function of nonmagnetic layer thickness $d_{NM}$ for Fe/Pt and Fe/W bilayers. (d) THz pulse width Δt as a function of $d_{NM}$ for Fe/Pt and Fe/W bilayers. (e) Extracted delay time $τ_D$ of the THz peak as a function of $d_{NM}$ for Fe/Pt and Fe/W bilayers.

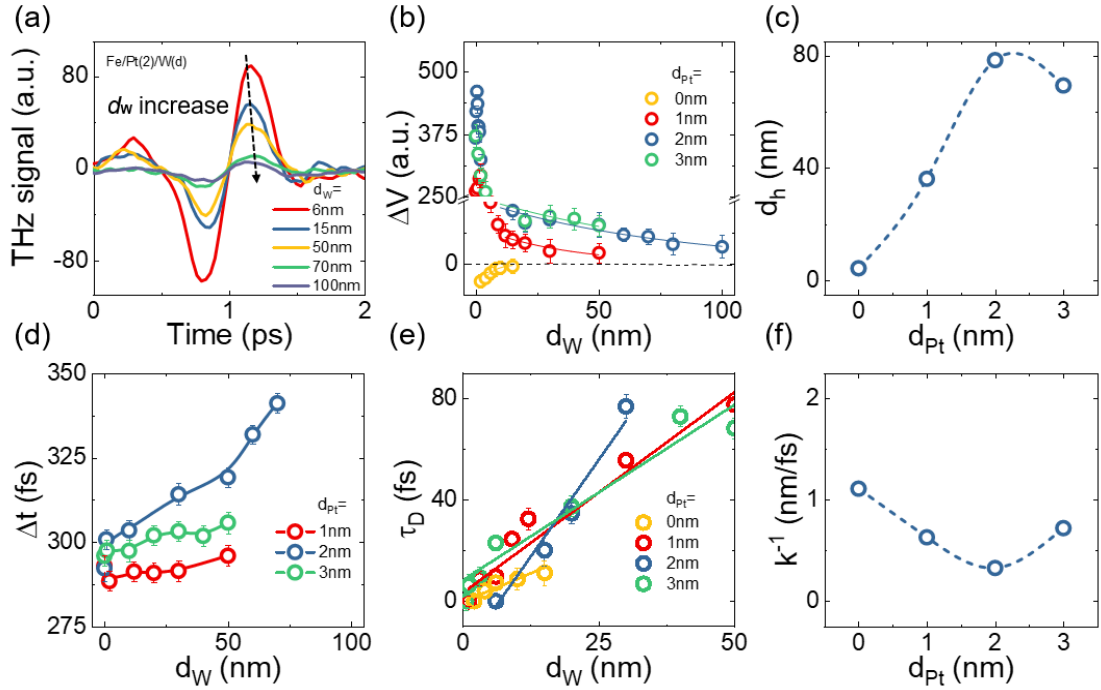

**Figure 3** (a) Time-domain THz waveforms of Fe(2 nm)/Pt(2 nm)/W($d_W$) trilayers with varying W thicknesses. (b) Extracted peak-to-peak THz amplitudes ΔV as a function of W thickness $d_W$ for different Pt thicknesses $d_{Pt}$. The solid lines represent fitted curves with an exponential formula. (c) Extracted decay length $d_h$ of the THz signal as a function of $d_{Pt}$. The dotted curve are guides to the eye. (d) THz pulse width Δt as a function of $d_W$ for different $d_{Pt}$. The solid curve are guides to the eye. (e) Delay time $\tau_D$ of the THz peak as a function of $d_W$ for different $d_{Pt}$, obtained from Hilbert transformation. (f) Transport velocity $k^{-1}$ extracted from the linear fitting of $\tau_D$ versus $d_W$ as a function of $d_{Pt}$. The dotted curve are guides to the eye.

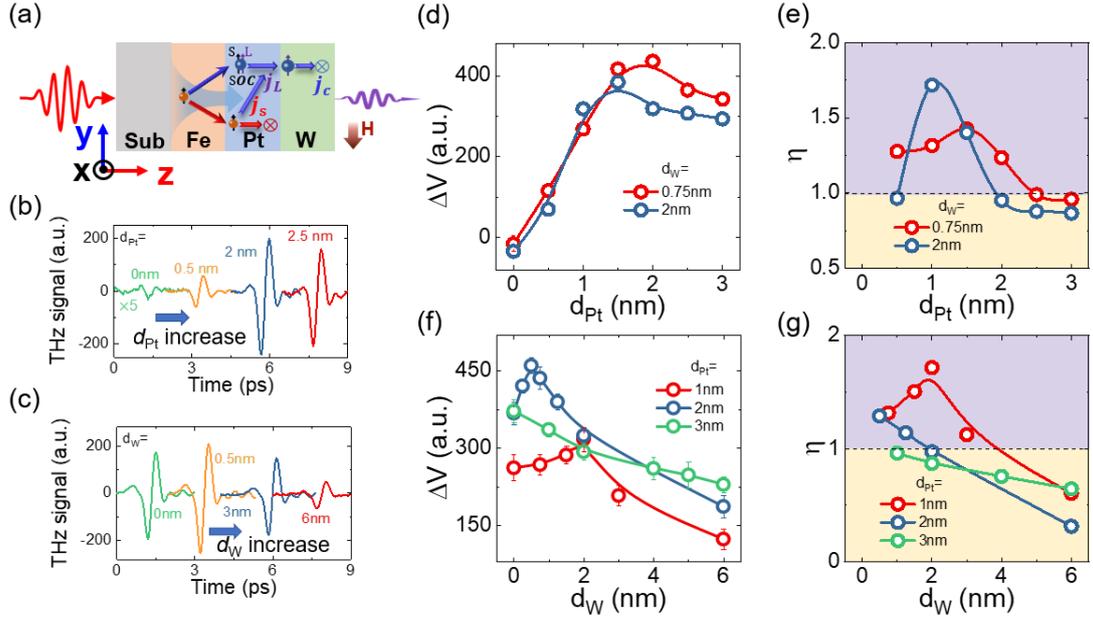

**Figure 4** (a) Schematic illustration of the Fe/Pt/W trilayer under femtosecond laser excitation. (b)-(c) Time-domain THz waveforms of (b) Fe/Pt($d_{Pt}$)/W(0.75 nm) with varying Pt thicknesses $d_{Pt}$, and (c) Fe/Pt(2 nm)/W($d_W$) with varying W thicknesses $d_W$. The curves are shifted horizontally for clarity. (d) Extracted peak-to-peak THz amplitudes ΔV as a function of $d_{Pt}$ for different W thicknesses. (e) Enhancement factor $\eta$ as a function of $d_{Pt}$ for different W thicknesses. (f) Extracted peak-to-peak THz amplitudes ΔV as a function of $d_W$ for different Pt thicknesses. (g) Enhancement factor $\eta$ as a function of $d_W$ for different Pt thicknesses.